\documentclass[manuscript]{aastex}

\shorttitle{A nanoflare model and the properties of coronal plasmas}
\shortauthors{L\'opez Fuentes \& Klimchuk}

\begin{document}

\title{A nanoflare based cellular automaton model and the observed properties of the coronal plasma}

\author{Marcelo L\'opez Fuentes\altaffilmark{1}}
\affil{Instituto de Astronom\'{\i}a y F\'{\i}sica del Espacio, CONICET-UBA, CC. 67, Suc. 28, 1428 Buenos Aires, Argentina}
\email{lopezf@iafe.uba.ar}

\and

\author{James A. Klimchuk}
\affil{NASA Goddard Space Flight Center, Code 671, Greenbelt, MD  20771, USA}

\altaffiltext{1}{Member of the Carrera del Investigador Cient\'{\i}fico, Consejo Nacional de Investigaciones Cient\'{\i}ficas y T\'ecnicas (CONICET), Argentina}

\begin{abstract}
We use the cellular automaton model described in L\'opez Fuentes \& Klimchuk (2015, ApJ, 799, 128) to study the evolution of coronal loop plasmas. The model, based on the idea of a critical misalignment angle in tangled magnetic fields, produces nanoflares of varying frequency with respect to the plasma cooling time. We compare the results of the model with active region (AR) observations obtained with the Hinode/XRT and SDO/AIA instruments. The comparison is based on the statistical properties of synthetic and observed loop lightcurves. Our results show that the model reproduces the main observational characteristics of the evolution of the plasma in AR coronal loops. The typical intensity fluctuations have an amplitude of 10 to 15\% both for the model and the observations. The sign of the skewness of the intensity distributions indicates the presence of cooling plasma in the loops. We also study the emission measure (EM) distribution predicted by the model and obtain slopes in log(EM) versus log(T) between 2.7 and 4.3, in agreement with published observational values.
\end{abstract}



\section{Introduction}
\label{intro}


Coronal loops are the basic observable blocks of the magnetically structured solar atmosphere. Any single theory proposed to explain the problem of coronal heating must be able to reproduce a series of diverse, and sometimes apparently contradictory, observed loop properties (see Klimchuk 2006, 2009; Reale 2010). Among other things, such theory should predict hot loops (2 to 4 MK) with apparently quasi-static evolutions, as observed in soft X-rays (Rosner et al. 1978), and highly dynamic cooler ($\approx$ 1 MK) loops that are too overdense to be in hydrostatic equilibrium, as observed in some EUV channels (Aschwanden et al. 2001). The multi-thermality of loops inferred from observations (Schmelz et al. 2009, and references therein) suggests that they might actually be structured in sub-resolution filaments or strands. It is possible then, that some of the observable properties of loops may be due to the collective contribution of thinner unresolved strands that follow more or less independent evolutions.


Regarding the heating process and consequent evolution of the plasma in the strands, one thoroughly considered possibility is that they are heated by short duration and small-scale impulsive events called nanoflares. This possibility corresponds to the scenario proposed during the 1980s by E. Parker (1988). The idea is that photospheric convective motions displace the strand footpoints, translating into magnetic stress between adjacent strands. When the stress due to misalignement between strands reaches a threshold, reconnection rapidly occurs releasing energy and heating the plasma. One or more of these reconnection events can combine to produce a nanoflare (see Klimchuk 2015). The presence of heating events like these applies naturally to impulsively evolving 1 MK loops. However, if nanoflares can occur at different frequencies the described mechanism could also explain quasi-static evolving hot loops. If the frequency is high enough the heating rate would be indistinguishable from a steady source (Cargill \& Klimchuk, 1997, 2004).


In search for a plasma diagnostic tool to discriminate the role of high versus low frequency heating, recent works have focused in the emission measure (EM) distribution of the plasma at different locations in active regions (AR), in particular in AR cores (see e.g., Bradshaw et al. 2012, Warren et al. 2012, Schmelz \& Pathak 2012, Reep et al. 2013). It has long been recognized that the left or ``colder'' side of the EM distributions follow a power law of the type $EM(T)~\propto~T^{\alpha}$ (Dere \& Mason 1993). The value of the $\alpha$ index can be used to estimate the relative contribution of the low and high frequency components of the heating. In the low frequency scenario the plasma in the strands is able to cool enough so that at any time there is a significant contribution of the colder plasma to the EM distribution. That would correspond to ``flatter'' distributions and smaller $\alpha$ values. On the other hand, if most of the plasma is in the high frequency heating regime the distribution is more peaked around the highest temperatures and $\alpha$ takes higher values. Through a series of hydrodynamic simulations, Bradshaw et al. (2012) found that the cut to determine the relative importance of the low frequency contribution is at $\alpha \approx 3$; above this threshold steady heating or high frequency nanoflares dominate. $\alpha$ index values obtained from observations in different wavelengths range from 2 to more than 4 (Tripathi et al. 2011, Schmelz \& Pathak 2012, Warren et al. 2011, Cargill 2014, Cargill et al. 2015). Warren et al. (2011) found that a simple mixture of 90\% high frequency and 10\% low frequency nanoflares could explain the observations at both high and low temperatures for many studied spectral lines. It has also been found that the relative contribution of both frequency components could be related to AR age (Ugarte-Urra \& Warren 2012).


In a preliminar theoretical paper that accompanies this one (L\'opez Fuentes \& Klimchuk 2015, henceforth Paper I) we study a cellular automaton model for the evolution of loop plasmas that is based in Parker's (1988) nanoflare scheme described above. In Paper I we analyze scaling laws of the plasma properties as a function of the parameters of the model and the presence of power laws in the nanoflare distribution. We provide a description of the model in Section~\ref{model}. In this paper we use the model to create synthetic loop lightcurves that we compare with real observations from Hinode/XRT and SDO/AIA. We base the comparison on the statistical properties of synthetic and observed lightcurves. We also obtain the power-law indices of the model's differential emission measure (DEM) distribution and compare it with observations from the works cited above.


The paper is organized as follows. In Section~\ref{observations} we describe the Hinode/XRT and SDO/AIA data used in the analysis. In Section~\ref{model} we provide a brief description of the model (a complete description can be found in Paper I). In Section~\ref{results} we describe our results: the statistical properties of modeled and observed lightcurves (Sub-section~\ref{lightcurves}) and the DEM distributions obtained from the model (Sub-section~\ref{DEM_slope}). Finally, we discuss and conclude in Section~\ref{conclusions}.


\section{Observations}
\label{observations}

\begin{figure*}[t]
\centering
\hspace{0.cm}
\includegraphics[bb=50 170 475 600,width=15cm]{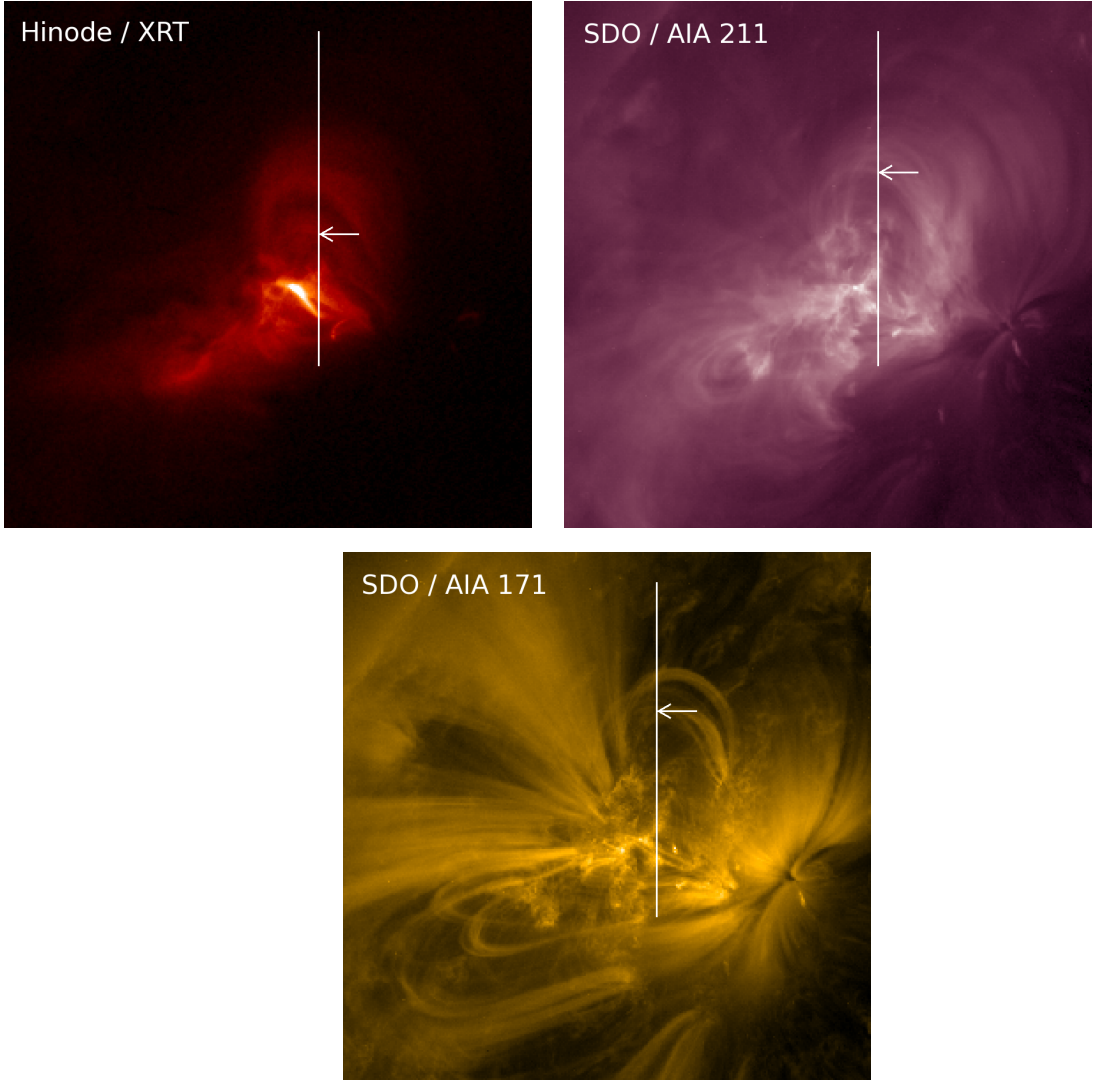}
       \caption{XRT, AIA 211 and AIA 171 images from the data sets used to obtain observed loop lightcurves. The time of the images is 18 January 2011, 11:54 UT, within a few seconds difference. This is the intermediate time of the sets. The vertical white lines correspond to the selected segments whose evolutions are shown in Figure~\ref{segments} and the arrows indicate the location of the loop structures whose lightcurves are presented in Figures~\ref{xrt_lc} to~\ref{aia171_lc}.}
\label{images}
\end{figure*}

We use simultaneous data from the X-ray Telescope (XRT, Golub et al. 2007) on board Hinode (Kosugi et al. 2007) and the Atmospheric Imaging Assembly (AIA, Lemen et al. 2012) on board the Solar Dynamics Observatory (SDO, Pesnell et al. 2012). The observation date is 2011 January 18 and the data correspond to active region NOAA 11147. The XRT observations have been obtained with the Al-poly filter and the AIA data correspond to the 171~\AA~and 211~\AA~bands (henceforth, AIA 171 and AIA 211). The observations begin at 10:49 UT and end at 12:59 UT. They cover a span of approximately 8000 s, with a time cadence of 10 s average for XRT and 12 s for AIA. For the analysis we construct data cubes containing 773 XRT images and 653 images for each of the AIA sets. We center each set on the AR of interest correcting for solar rotation and then we carefully coalign the three sets. In Figure~\ref{images} we show examples of coaligned images from each of the sets for an intermediate observation time (11:54 UT). The structures observed in the three wavelengths show clear differences.

To obtain lightcurves from coronal loop pixels we select corresponding vertical (north-south) segments in the three datasets (indicated with white lines in the three panels of Figure~\ref{images}) and we study the intensity evolution of the pixels along the segments by arranging them in rectangular arrays with the time corresponding to the horizontal axis, as shown in Figure~\ref{segments}. From these arrays we identify and select horizontal rows that correspond to the evolution of bright pixels from coronal loop structures. We choose  different pixels and therefore different loops for the three observing channels, as explained later. The upper-left panels in Figures~\ref{xrt_lc} to~\ref{aia171_lc} show loop lightcurves obtained in this way. The corresponding loops and lightcurve locations are indicated with arrows in the panels of Figures~\ref{images} and~\ref{segments}.

The prominent bright feature near the lower central part of the arrays in Figure~\ref{segments} (around XRT pixel index 40), corresponds to a small flare at the core of the AR that ocurred between 11:35 and 11:50 UT. The spatial and temporal location of this event in the three segment arrays confirms the correct coalignment of the datasets. Since in this work we are interested in the evolution of non-flaring loop structures, we do not include in our study lightcurves crossing the site of the small flare. Comparing the XRT and AIA segment arrays in Figure~\ref{segments} it can be seen that, besides the flare and the clear bottom-up decrease of intensity (going farther from the AR core) in all arrays, no obvious visual correlation can be appreciated between the evolution of XRT and AIA 171 and 211 structures. However, since we do not perform a detailed analysis in search of possible correlations, we cannot discard their presence. Other studies based on the quantitative analysis of time-lags between lightcurves from different wavelengths have systematically found such correlations (see Viall \& Klimchuk, 2012, 2013 and 2015).


\section{Implementation of the model}
\label{model}

We use the 2D cellular automaton (CA) model that is thoroughly described in Paper I. A 1D model based in the same ideas has been studied in L\'opez Fuentes \& Klimchuk (2010). The present model is a simplified 2D version of the 3D scenario envisioned by Parker (1988), in which magnetic strands are represented as vertical structures connecting two distant portions of the photosphere represented by horizontal planes. As the strand footpoints are randomly displaced by photospheric motions, strands tangle and the magnetic stress between interacting strands creates the conditions for energy release by reconnection. In our model, magnetic strands are represented by moving elements in a 2D grid, which are randomly displaced simulating photospheric motions. These moving elements play two roles: 1) displacements are associated with the increasing inclination (and magnetic stress) of the strands, and 2) as the elements travel across the grid they encounter other elements with which they interact simulating the existence of points of contact between tangled strands. Whenever the simulated mutual inclination of interacting strands excedes a critical value, they reconnect and energy from the magnetic stress is released to heat the plasma. All the reconnection events that a strand suffers in a given time step (rarely more than two) are combined in a single nanoflare. The output of the model is the set of nanoflare energies released in each strand during the evolution. Despite of its simplicity, the model incorporates the basic characteristics of the physical process behind Parker's proposal. The main motivation for this simplicity is the ability to numerically deal with a large number of runs in reasonable computing times.

The input parameters of the model are the strand length ($L$), the vertical magnetic field ($B_v$), the critical misalignment angle ($\theta_c$), the number of strands ($N$), the photospheric displacement of the footpoints during each time step ($d$), the time step duration ($\delta t$), and the nanoflare duration ($\tau$). We fix some of these parameters to typical solar values and vary the rest of them, one at a time, to obtain different model outputs to be compared with the observations. Parameters like $d$ and $\delta t$ are set to 1000 km and 1000 s for all the computations, because these are the typical scales of the size and turnover time of photospheric granules (see e.g., Foukal 2004).

Although in previous studies we considered $L$ and $B_v$ to be independent parameters, there is known to be an overall correlation between length and field strength. We therefore use the relationship found by Mandrini et al. (2000): $B \propto L^{-0.88}$. Using a representative value of the proportionality constant (courtesy of C.H. Mandrini, private communications) we find that, for $L$ expressed in Mm,

\begin{equation}
\label{b_lock}
B_v \approx 10^4 \times L^{-0.88} \mbox{G}.
\end{equation}

\noindent As analyzed in Paper I, from this relation loop lengths in the range 40-120 Mm correspond to magnetic strengths of approximately 100-400 G. These values are completely reasonable for solar standards. In general, we expect the field strength to vary along the strands as the flux tubes expand (see e.g., Warren \& Winebarger 2006). The field strength in the Mandrini et al. expression refers to the average along the strand, and so Equation~\ref{b_lock} is appropriate for our simple model in which all strands have a constant width. It is worth noting that distinct coronal loops seem to have a roughly constant width (Klimchuk 2000; L\'opez Fuentes et al. 2006).

In what follows we use Equation~\ref{b_lock} to lock the magnetic strength parameter to the loop length. This leaves us with the free parameters: $L$, $\theta_c$, $N$ and $\tau$. The effect that these parameters have on the model output has been extensively studied in Paper I. We use some of these results in our analysis below.

Since our main objective is to compare the model with observations, we need to generate from the CA output a simulated plasma response and the corresponding observable emission. The output of the CA model consists of the energies of a series of nanoflares acting on each of the $N$ strands that comprise the model. We obtain the density and temperature evolution of the plasma on each strand using the EBTEL code (Klimchuk et al. 2008, Cargill et al. 2012), which is particularly useful for the large number of computations involved here. Given the number of events and strands to be computed, a regular 1D hydrodynamic code would be numerically too costly.

From the evolution of the computed plasma number density ($n$) and temperature ($T$) we use the known instrument response, $S(T)$, to compute the observed intensity, $I$, using the relation:

\begin{equation}
\label{response_eqn}
I = \int S(T)~n^2~dV.
\end{equation}

\noindent From this expression we compute the contribution of each strand to the intensity per pixel observed by each instrument. In Figure~\ref{response} we show a log-log plot of the instrument response functions versus temperature for the three observation channels analyzed here: $S_{XRT}(T)$ (dotted line), $S_{AIA211}(T)$ (dot-dashed line), and $S_{AIA171}(T)$ (dashed line). The plot illustrates the motivation behind the channel choice, since they provide a reasonably continuous temperature coverage (i.e., AIA 211 fills the temperature gap between the XRT and AIA 171).

Figure~\ref{single_strand} illustrates the process of obtaining the synthetic emission produced by a single strand. The top panel shows a set of consecutive nanoflares that heat the strand (only a small portion of the full model evolution is shown), the next two panels correspond to the temperature and density evolution computed by the EBTEL model, and finally, the bottom panel shows the computed XRT and AIA 171 and 211 intensity evolutions obtained from Equation~\ref{response_eqn}. Adding the emission of all the strands covered by a single pixel we construct synthetic lightcurves for each instrument. We simulate the photon noise present in the observational measurements by adding a Poisson distribution of the amplitude indicated by XRT and AIA instrument calibrations (Narukage et al. 2011, Boerner et al. 2012). In the case of AIA data the loop emission is typically less than half of the total emission in a pixel (Viall \& Klimchuk 2011), so we add a background intensity to the synthetic lightcurves. Based on the minimum observed intensities of the segment arrays from Figure~\ref{segments}, we use 400 DN pix$^{-1}$ s$^{-1}$ for AIA 171 and 700 DN pix$^{-1}$ s$^{-1}$ for AIA 211. These values are also similar to the mean intensities of the respective data cubes.

In order to find what combination of model parameters best reproduces the observed lightcurves, we consider the following. As we describe above, we lock the magnetic field parameter, $B_v$, to the loop length, $L$, using Equation~\ref{b_lock}, and therefore any change in $L$ implies an automatic change in $B_v$ following that expression. From the results from Paper I we know that: 1) an increase of $L$ produces a decrease of the lightcurve intensities, 2) an increase of $\tan \theta_c$ increases the intensities, 3) an increase of $N$ or $\tau$ smooths the model lightcurve fluctuations. These parameter dependencies and the corresponding scalings are studied in detail in Paper I. Although $\theta_c$ is uncertain, we set $\tan \theta_c$ = 0.25, based on Parker's (1988) argument that equates the observed energy losses from the corona with the stressing of the coronal magnetic field by observed photospheric driving, and based on the onset of secondary instability studied by Dahlburg et al. (2005, 2009). Then, we adjust the intensity of the synthetic lightcurves using just $L$ (or rather the $L$ and $B_v$ combination) as a free parameter. Since $N$ and $\tau$ have similar effects on the fluctuations, we fix $\tau$ = 200 s and use $N$ as the free parameter for the amplitude of the intensity fluctuations. For the diameter of the model loop that contains the $N$ strands we use a conservative value of 3 Mm, that corresponds to the typical diameter of XRT loops. Therefore, increasing $N$ implies strands with smaller cross sections, so the intensity of each strand is smaller but the total loop intensity remains approximately unchanged. Clearly, an increase of the number of strands tends to smooth the lightcurve of the combined strands, and that explains its effect on the fluctuations.

In summary, we vary $L$ until we find a good match of the mean intensity of the synthetic lightcurves to the observations, and then we vary $N$ until the standard deviation of the fluctuations coincides with the observed cases. Although the choice of $L$ and $N$ as the only adjustment parameters may sound rather arbitrary, it should be regarded just as a practical measure considering that our prime objective is to study the plausibility of the present model to reproduce the observed coronal plasma evolution. We expect that further refinement of the parameters will be possible in the future.


\section{Results}
\label{results}

\subsection{Observed and model lightcurve comparison}
\label{lightcurves}

\begin{figure*}[t]
\centering
\hspace{0.cm}
\includegraphics[bb=55 60 560 730,width=14cm]{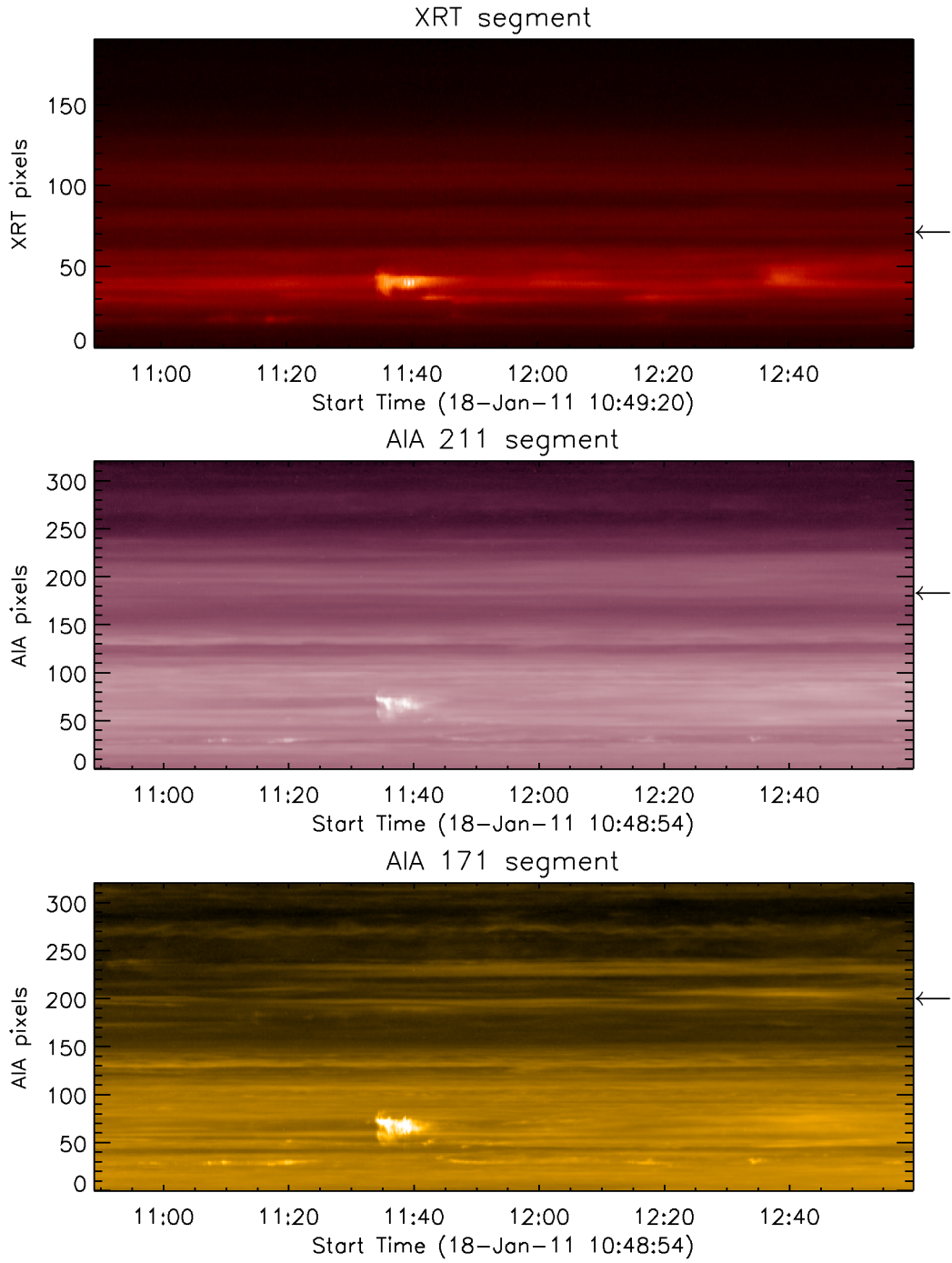}
       \caption{Evolution of the pixel segments indicated with white vertical lines in Figure~\ref{images} for the full timespan of the datasets consisting of 773 XRT and 653 AIA images.}
\label{segments}
\end{figure*}

\begin{figure*}[t]
\centering
\hspace{0.cm}
\includegraphics[bb=70 365 550 700,width=14cm]{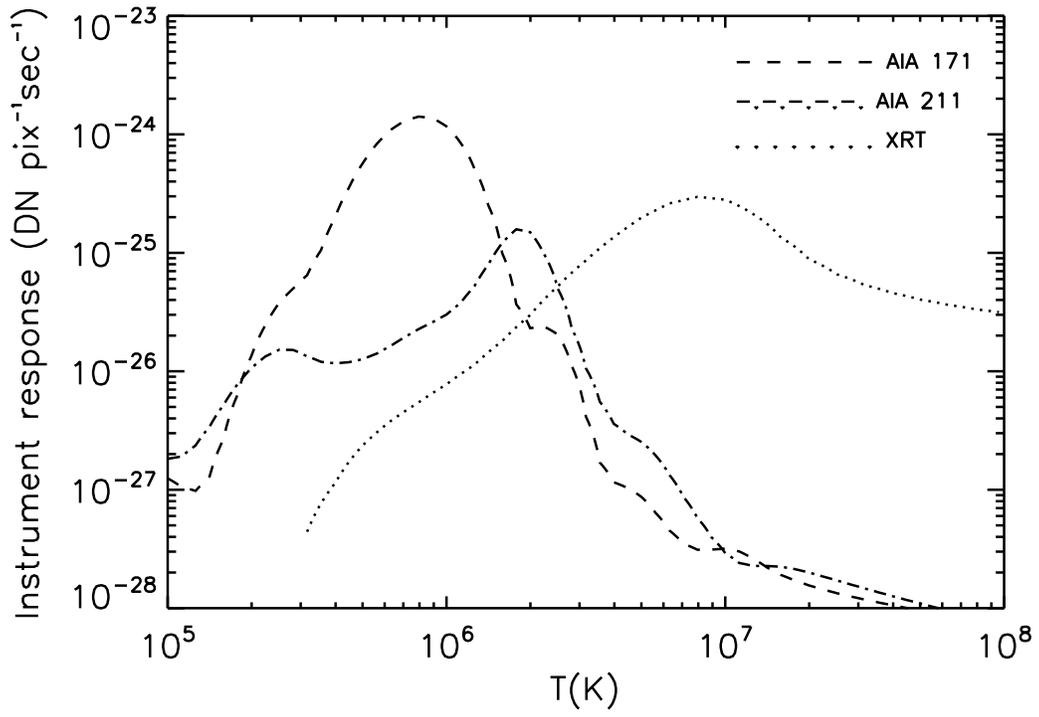}
       \caption{XRT (Al-poly filter), AIA 211 and 171 instrument responses versus temperature. Notice that AIA 211 tends to fill the temperature gap between XRT and AIA 171.}
\label{response}
\end{figure*}

We apply the procedure described in Section~\ref{model} to create synthetic signals that reproduce the statistical properties of selected observed lightcurves from the segment evolutions shown in Figure~\ref{segments}. As we discuss below, in its fully developed regime (i.e., when all strands are interacting and producing nanoflares) the signal produced by the model presents fluctuations around a certain mean intensity level that is maintained during the evolution. Therefore, for a reasonable comparison with the model we select observed lightcurves that do not present abrupt changes of the mean level of intensity or long term trends that can be due to processes beyond the applicability of the model (i.e., large scale flaring or changes due to the evolution of the photospheric magnetic flux). Given the random nature of the problem we obviously do not expect to find model lightcurves that match the observations point to point. We can only base the comparison on the statistical properties of the synthetic and observed lightcurves. We find that reasonable solar values for the model parameters can produce lightcurves with statistical properties similar to those of observed lightcurves.

The observed lightcurves chosen in the different channels do not come from the same corresponding pixels. The reason is as follows. We are attempting to study distinct loop structures, but such structures produce only part of the total emission observed along the line-of-sight in a pixel. It can be a minor fraction for 211 and 171. In those cases, most of the emission comes from a diffuse component, sometimes called the background. Since we make no attempt to model the diffuse component, and since its contribution to the total emission varies from channel to channel and from pixel to pixel, we decided to choose separate pixels for each channel based on a relatively strong loop contribution.

In Tables~\ref{xrt_table} to~\ref{aia171_table} we show the results of the observed and modeled lightcurves comparison. In these tables, each subdivision contains the statistical properties of the observed (upper row) and corresponding modeled (lower row) lightcurves. The first column provides the position, in instrument pixels, of the observed lightcurve in the corresponding array of Figure~\ref{segments}. The last column indicates the model parameters ($L$, $N$) used in the simulations. We run CA models with 200 time steps, therefore the synthetic lightcurves have a duration of 2$\times 10^5$ s. The statistical properties computed from the observed and synthetic intensities are: mean, median, standard deviation, ratio of standard deviation to mean, and skewness. In the tables we include observed lightcurves from a variety of locations in the selected segments that satisfy the above conditions (no flaring or obvious long term variation). It can be seen that the synthetic lightcurves reproduce the statistical properties of the observed cases (except in the cases discussed below) for reasonable values of the model parameters. Once again, given the random nature of the modeled system, an exact match of the properties is not expected. Even running the codes twice with exactly the same parameters, produces small differences in the mean, median and standard deviation.

In Figures~\ref{xrt_lc} to~\ref{aia171_lc} we show comparable observed and synthetic lightcurves of the same durations ($\sim$ 8000 s). In these cases we select, from the full synthetic lightcurves of 2$\times 10^5$ s durations, portions of 8000 s that have the same statistical properties and that visually resemble the observed cases. In Figures~\ref{xrt_lc} to~\ref{aia171_lc} the upper-left panels correspond to the observed lightcurves indicated with arrows in the corresponding panels of Figure~\ref{images} and the evolution arrays of Figure~\ref{segments}. The indices (pixel positions) of these observed lightcurves in the corresponding panels of Figure~\ref{segments} are: 72 for XRT, 183 for AIA 211, and 199 for AIA 171. Notice that these observed lightcurves are included in Tables~\ref{xrt_table} to~\ref{aia171_table}. The set of parameters used to obtain the XRT synthetic lightcurve of Figure~\ref{xrt_lc} upper right panel are: $L$ = 88 Mm and $N$ = 121. For the AIA 211 lightcurve of Figure~\ref{aia211_lc} upper-right panel we use: $L$ = 80 Mm and $N$ = 49. For the AIA 171 synthetic lightcurve of Figure~\ref{aia171_lc} the parameters are: $L$ = 85 Mm and $N$ = 49. In the lower panels of Figures~\ref{xrt_lc} to~\ref{aia171_lc} we plot the histograms of the corresponding observed and synthetic lightcurves. The statistical information is provided in the text insets in the panels. These statistical properties are consistent with the results from Tables~\ref{xrt_table} to~\ref{aia171_table}.

In nearly all cases, our model can reproduce with reasonable solar parameters the observed basic statistical properties (mean intensity and fluctuation sizes) of the coronal emission for three different wavelength bands. The exceptions are the three high-intensity AIA 171 cases shown in the top rows of Table~\ref{aia171_table}, in which the model that reproduces the right intensity level has fluctuations that are too large compared to the observations. To reduce the relative size of these fluctuations, we increase $N$ to 441, but it is not enough to smooth the model lightcurves. Further increasing $N$ does not improve much the results. For the lower intensity cases of the two last rows, the model successfully reproduces both the mean intensity and the standard deviation. One possibility for this inconsistency is that the background level (400 DN pix$^{-1}$ s$^{-1}$) that we are using to simulate the lightcurves is not the right one for the high intensity cases. We might be underestimating the background at those locations. From the panels of Figure~\ref{segments} it is clear that the intensity decreases as one goes up in the segments (moving farther from the AR core). It is possible that 400 DN pix$^{-1}$ s$^{-1}$is a reasonable background for the low intensity cases but it is too small for intensities above 1000 DN pix$^{-1}$ s$^{-1}$. If the real background is higher, the relative size of the fluctuations provided by the model will be relatively smaller. Notice that in the cases of the last row of Table~\ref{aia211_lc} and the last two rows of Table~\ref{aia171_lc} the loop intrinsic emission is only a fraction of the background level, as is frequently the case for loops observed in some EUV channels (see e.g., Aschwanden et al. 2007).

In Tables~\ref{xrt_table} to~\ref{aia171_table} we listed the lightcurve indices in increasing order, so the mean intensity becomes ordered in approximate decreasing order (as one moves upward along the segments and farther away from the AR core). As we discussed in Section~\ref{model}, the intensity of the model lightcurves decreases as the loop length parameter, $L$, increases. This can also be seen in Tables~\ref{xrt_table} to~\ref{aia171_table} and is consistent with the fact that we are modeling loops that are progressively longer as we move away from the AR core. The $L$ parameter values that give the best agreement between models and observations are completely reasonable for the location of the analyzed lightcurves within the AR. Typical measured lengths of the clusters of loops covered by the segment evolutions shown in Figure 2 are in the range $\sim$70-100 Mm for XRT observations and $\sim$40-100 Mm for AIA observations. These values correspond, of course, to the projection of the loops on the plane of the sky (i.e., as observed in the images of Figure~\ref{images}), however, given the position and orientation of the loops in the AR we consider that they do not differ much from the actual loop lengths.

Tables ~\ref{xrt_table} to~\ref{aia171_table} can help us understand the sensitivity of the model intensity to variations in $L$. We start by comparing the first two cases in each table, because these are the brightest and therefore least affected by the assumed background. For the XRT loop models with $L=73$ Mm and $L=77$ Mm (Table~\ref{xrt_table}), we see that a 5\% increase in length produces a 17\% decrease in intensity. For AIA 211, an increase of 8\% in length between the loops with $L=42$ Mm and $L=47$ Mm (Table~\ref{aia211_table} ) produces a decrease of 22\% in intensity. A length increase of 10\% between AIA 171 loops with $L=46$ Mm and $L=51$ Mm (Table~\ref{aia171_table}) produces an intensity decrease of 23\%. Thus, the magnitude of intensity decrease is roughly 3 times the magnitude of length increase. The influence of the 400 DN pix$^{-1}$ s$^{-1}$ background can be seen in the two AIA 171 low intensity cases. A 6\% length increase from $L=80$ Mm to $L=85$ Mm (Table~\ref{aia171_table}) corresponds to a comparable magnitude intensity decrease of 7\%. If we subtract the background, the change in the intrinsic loop intensity provided by the model increases to 20\%, which follows the trend for bright loops. These examples show that the model output intensity is clearly sensitive to variations in length. The best fit models in Tables~\ref{xrt_table} to~\ref{aia171_table} are those that most closely reproduce the observed mean intensity, where we consider length increments of 1 Mm. In one extreme case, a 1 Mm length difference produced a 7\% intensity difference.

From Figures~\ref{xrt_lc} to~\ref{aia171_lc} it is interesting to note that there are two components of the fluctuations: a short-term/small-amplitude contribution and another longer-term/larger-amplitude contribution. In the case of the model lightcurves we know that the first contribution is due to the simulated photon noise. We assume that this is also the case for the observations. One way to identify the short term fluctuations is to compute the root mean square deviations with respect to the 10 point running average of the signal. These variations are of the order of 2 to 4\% of the mean intensity and coincide with the photon noise expected amplitude. The size of these fluctuations is only a fraction of the approximate 10-15\% of the typical standard deviations shown in Tables~\ref{xrt_table} to~\ref{aia171_table}. Therefore, we conclude that the longer term fluctuations that dominate the measured standard deviation are due to intrinsic variations of the coronal plasma emission. In the case of the synthetic lightcurves, we know that the larger amplitude fluctuations are due to the individual evolutions of the different strands that compose the loop. The comparison suggests that this might also be the case for the observed loops.

It is possible to extend the statistical comparison between observed and synthetic lightcurves beyond the mean and the standard deviation. For instance, the degree of asymmetry of the intensity distribution can provide important information on the mechanisms behind the evolution of the emitting plasma. Terzo et al. (2011) suggested that the difference between the mean and the median of the intensity distributions in XRT observations may be associated with the widespread presence of cooling plasmas in the corona. Here, we use the skewness parameter of the distributions as a measure of the asymmetry. A positive skewness indicates that the events located to the left of the mean value are less spread than those to the right, so that the right ``tail'' is longer and less prominent than the left one. The opposite is true for negative skewness. The shapes of the distributions shown in the lower panels of Figures~\ref{xrt_lc} to~\ref{aia171_lc} suggest that they all have positive skewness. This is confirmed by the computed values indicated in insets at the top-right of the lower panels and in all the observed and modeled lightcurves listed in Tables~\ref{xrt_table} to~\ref{aia171_table}. This result is consistent with the finding of Terzo et al. (2011) that the median intensity is less than the mean.

It is worth mentioning that the skewness of the observed and model lightcurves shown in Tables~\ref{xrt_table} to~\ref{aia171_table} and Figures~\ref{xrt_lc} to~\ref{aia171_lc} have the same sign (positive) and are of the same order, but they do not exactly agree. We explored whether there is a dependence of the skewness on the model parameters. We did not find any such dependence. In different subsets of long simulations lasting up to 10$^6$ s, the amplitude of skewness changes, but the sign is always positive. Therefore, we conclude that although the sign of the skewness provides an indication of the kind of processes at play in the plasma evolution, its amplitude is not related to the parameters of our model. As mentioned above, Terzo et al. (2011) found that the median is systematically less than the mean in intensity distributions observed by XRT. We find the same result for the overwhelming majority of observed and modeled light curves in Tables~\ref{xrt_table} and~\ref{aia171_table} (second and third columns). We fully agree with their conclusion that this is consistent with impulsive heating, since the heated coronal plasma spends most of the time in the cooling phase of its evolution. This also agrees with the time-lag measurements of Viall and Klimchuk (2012, 2013, 2015), and the recent work by Tajfirouze et al. (2015) based on the modeling of observed loop lightcurves. We note that, as in the case of the skewness, we could not find a systematic relation between the numeric difference of the mean and  median and the parameters of the model.


\subsection{Emission measure distribution slope}
\label{DEM_slope}

To further compare the CA-EBTEL model combination with observations, we analyze here the emission measure (EM) distribution of the plasma. As discussed in Section~\ref{intro}, in a series of recent works (see e.g., Bradshaw et al. 2012, Schmelz \& Pathak 2012, Warren et al. 2012) authors began to study the slope of the distribution in log(EM) versus log(T) plots, as a diagnostic tool for the evolution of the observed plasma. The presence of a constant slope for a wide range of temperatures implies a scaling-law of the type EM~$\propto$~T$^\alpha$. We will call $\alpha$ the slope index. It has been argued that since this slope determines the amount of plasma that reaches low temperatures, it can be used as an indicator of the proportion of the contribution of high versus low frequency heating.

Here, we analyze the differential emission measure distribution DEM provided by the EBTEL model for a single strand from the CA. Since DEM = EM/T, the DEM slope is one less than the EM slope. We run the model for different values of $L$ using a time resolution of 1 s and a total evolution of 2$\times10^5$ s. Therefore, the DEM distribution contains 2$\times10^5$ thermal-state ``samples'' of the plasma. The magnetic field parameter, $B_v$, is set according to Equation~\ref{b_lock} and the rest of the parameters are fixed in $\theta_c$ = 0.25, $N$ = 49, $\tau$ = 200 s. The slopes of the distributions are computed in the range of temperatures between 1 MK and the temperature of peak emission measure. As an example, in Figure~\ref{dem} we plot log(DEM) versus log(T) for a run with $L$ = 100 Mm. In this case, the DEM peaks at 4 MK and the slope is 2.47. The corresponding EM slope is 3.47.

For each value of $L$ we find that different runs provide slightly different values of the slopes. This is due to the randomness of the nanoflares produced by the model and the consequent small differences in the temperature distributions produced. To account for these small differences we run the codes 5 times for each value of $L$ and we compute the average DEM slope. In Table~\ref{slope_indexes} we show the results of each of the 5 runs for each value of loop length. The mean is given at the bottom. To compare with observational works based on EM distributions, we add 1 to the DEM slopes and obtain mean EM slopes ranging  between 2.7 and 4.3. This agrees with the published observational values.

The EM slopes in Table~\ref{slope_indexes} are based on individual strands. However, we sample the strand 2$\times10^5$ times (once each second) during its evolution to obtain a time average. As long as the $N$ strands of a given simulation have similar time-averaged properties, the result for a single strand should be equivalent to that from the full collection of strands. This is true even if the full collection is averaged over much shorter time intervals, such as the tens of seconds of a typical exposure used to obtain EUV spectra from which EM distributions are derived. We have verified that the EM slopes of different strands from the same simulation are very similar. A relevant point is that, although the evolution and heating of each strand depends on the evolution and heating of other strands, the model does not produce strong collective behavior. Localized bursts of multiple reconnection events do not occur. If they did, the time average of a single strand would not necessarily be a good representation of a snapshot of many strands.

We have verified the individual strand approach by computing the combined DEM from all of the strands in an $L=$ 100 Mm, $N=$ 225 run in several different 30 s windows, corresponding to a typical Hinode/EIS integration time. An example is shown in Figure~\ref{obs_dem}. We have binned the DEM over log$T=$ 0.1 intervals, as is commonly done with actual observations. The slope of 2.51 is in good agreement with the values given in Table 4. The slopes for the other windows are also consistent with the table, though there is a larger spread due to fact that the DEM curves are noisier than those from the 2$\times10^5$ s single strand averages, as expected.
 
In Figure~\ref{slope_length} we plot the mean slope indices from Table~\ref{slope_indexes} versus $L$. Interestingly, the plot follows a linear trend. A linear regression provides a slope of 0.014 as indicated in the panel of the Figure. This implies that the model predicts a dependence of the DEM slope index with the loop length of the type $\alpha \propto L$. We discuss the origin of this dependence in Section~\ref{conclusions}.


\section{Conclusions}
\label{conclusions}


In this paper we compare coronal observations from the X-ray Telescope (XRT) on board Hinode and the Atmospheric Imaging Assembly (AIA) on board SDO with synthetic data created using a Cellular Automaton (CA) model studied in an accompanying paper (L\'opez Fuentes \& Klimchuk 2015, Paper I). The model is based on the idea that loops are made of elementary strands that are shuffled and tangled by photospheric motions at their foopoints, producing magnetic stress which is released in the form of reconnection and plasma heating (Parker 1988). The output of the model is a series of nanoflares that heat the different strands in the loop. We compute the response of the plasma to this heating using the hydrodynamic code Enthalpy Based Thermal Evolution of Loops (EBTEL, Klimchuk et al. 2008, Cargill et al. 2012). From the known response of the instruments and the temperature and density obtained with EBTEL we compute synthetic lightcurves that we compare to the observations. We selected lightcurves from a series of loops observed in AR 11147 on 2011 January 18 with Hinode/XRT and with SDO/AIA in the 211~\AA~and 171~\AA~channels. For the comparison we computed the main statistical properties of both observed and synthetic lightcurves. Our results show that using reasonable solar parameters the model can reproduce the statistical properties of observed lightcurves.


One of the studied statistical properties is the standard deviation, which we use as an indicator of the relative amplitude of the signal fluctuations. We found that the typical fluctuations have an amplitude of 10 to 15\% both for the model and the observations. This is also consistent with observations reported by other authors (see e.g., Warren et al. 2010). From Figures~\ref{xrt_lc} to~\ref{aia171_lc} it can be seen that the typical duration of the fluctuations in both observed and modeled lightcurves is around 1000 s. Although the temporal span of the observations is too short to perform a reliable frequency analysis, doing this in the case of the synthetic lightcurves confirms that this is the main frequency of the fluctuations. This is not a surprising result, since 1000 s is the duration of the CA model time step, which corresponds to the typical turnover time of photospheric granules. It was also found in Paper I that the typical waiting times between consecutive nanoflares is one or two time steps. The fluctuation in the synthetic lightcurves is due to the superposition of the emission evolution from different strands, so there must be some degree of coherence among the strands. The similarity of the durations in the observed and modeled lightcurves shown in Figures~\ref{xrt_lc} to~\ref{aia171_lc} suggests that that could also be the case for real loops. Similar durations have been reported recently by Ugarte-Urra \& Warren (2014).


The shape of the intensity distribution can provide information about how the plasma in the loops evolves. In search of asymmetries in the distributions, here we compute the skewness and we found that in all the selected observed lightcurves and the corresponding modeled cases the skewness is positive. This indicates that the right ``tail'' of the distributions is more spread and less bulky than the left one. This means that there is a larger weight of the low intensity part of the fluctuations. In other words, there are more intensity counts below the mean of the distribution than above it. Terzo et al. (2011) argued that this is related to the widespread presence of cooling plasma in the loops. As a proxy of the distribution asymmetries they used the difference between the mean and the median. They found that the median is systematically smaller than the mean. Here we compute both the mean and the median for our observed and modeled cases and confirmed that result. A very recent work by Tajfirouze et al. (2015), also based on observed lightcurves modeling, confirms the widespread presence of hot plasma cooling.

\begin{figure*}[t]
\centering
\hspace{0.cm}
\includegraphics[bb=54 360 564 1011,width=14cm]{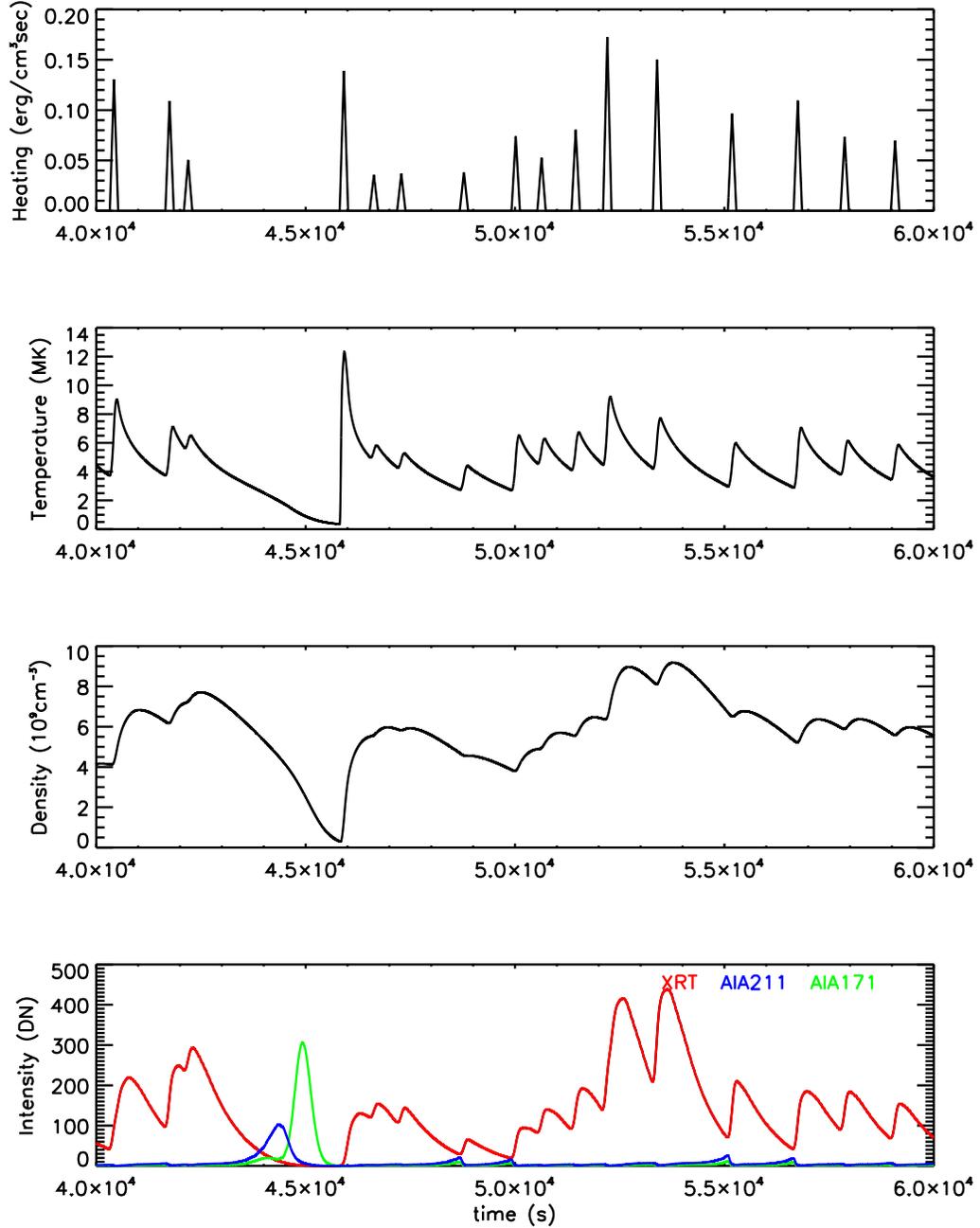}
       \caption{Portion of the evolution of a single strand from the CA model. Top panel: set of triangular nanoflares from the model. Second and third panels are the corresponding temperature and density of the plasma obtained with the EBTEL model. The lower panel shows the observable emission detected by XRT (red), AIA 211 (blue) and AIA 171 (green) computed using Equation~\ref{response_eqn} and the known instrumental response $S$($T$).}
\label{single_strand}
\end{figure*}
\begin{figure*}[t]
\centering
\hspace{0.cm}
\includegraphics[bb=60 200 580 600,width=17.cm]{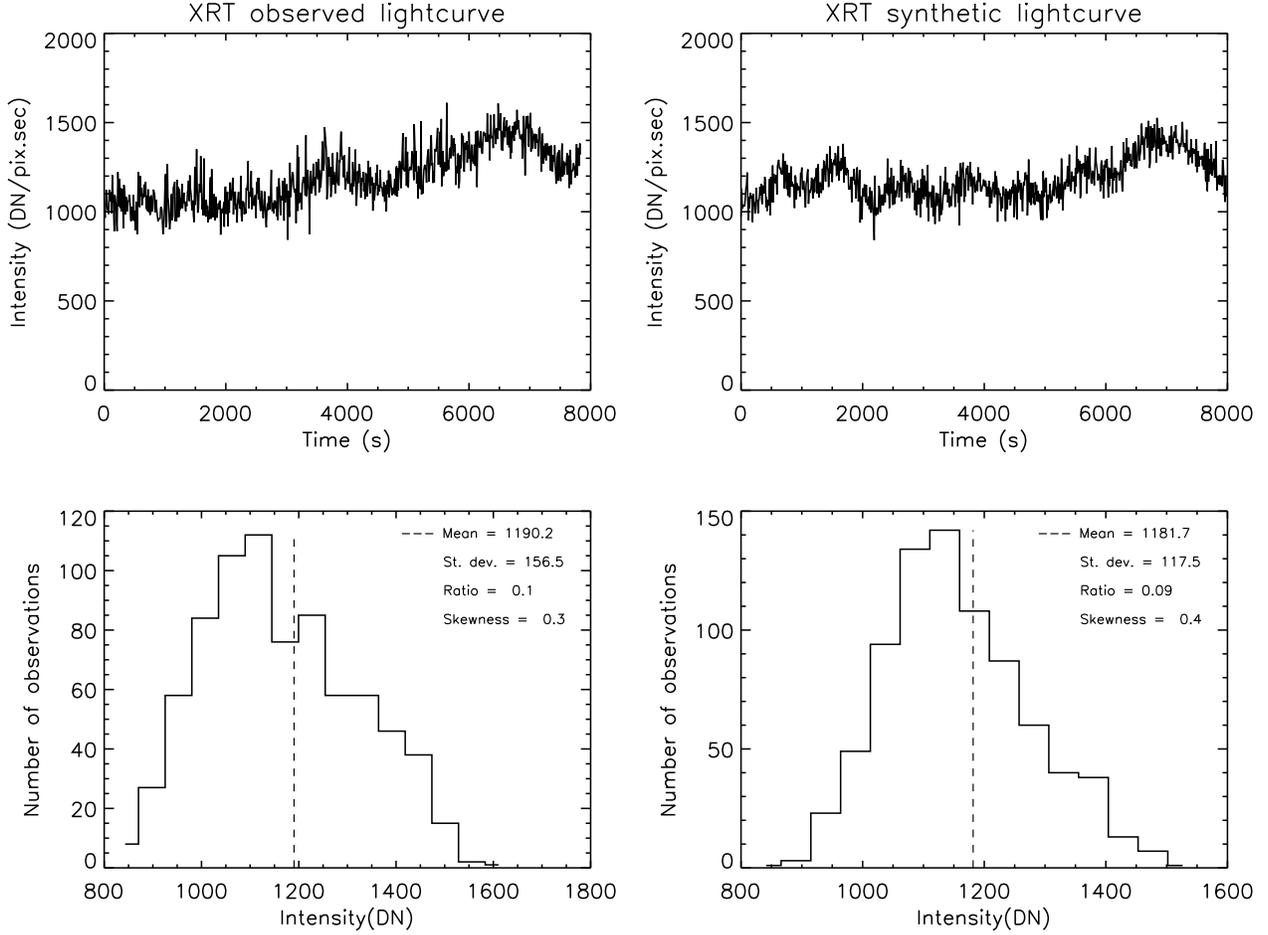}
       \caption{Similar observed (upper-left panel) and synthetic (upper-right panel) XRT lightcurves. The observed XRT lightcurve corresponds to the location indicated by arrows in the corresponding panels of Figures~\ref{images} and~\ref{segments}. The parameters used for the model are $L=88$ Mm and $N=121$. In the lower-row panels the histograms of the upper-row lightcurves are shown. The statistical properties: mean, standard deviation, ratio of standard deviation to the mean and skewness, are provided in the text insets of the panels. The model reproduces the main statistical properties of the observed lightcurve.}
\label{xrt_lc}
\end{figure*}
\begin{figure*}[t]
\centering
\hspace{0.cm}
\includegraphics[bb=60 200 580 600,width=17.cm]{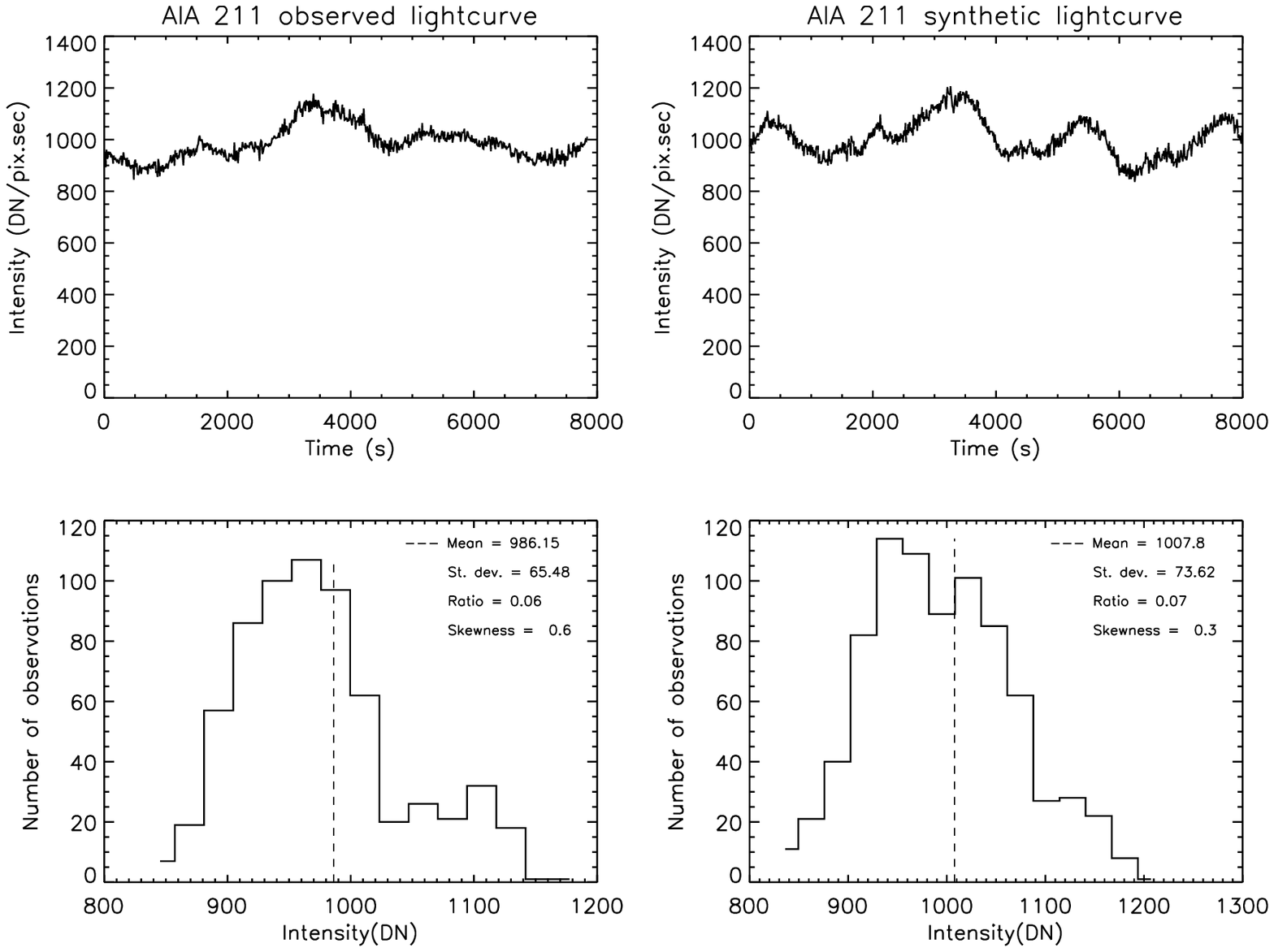}
       \caption{Idem Figure~\ref{xrt_lc} for AIA 211. The observed lightcurve corresponds to the location indicated with arrows in the corresponding panels of Figures~\ref{images} and~\ref{segments}. The parameters used for the model are $L=80$ Mm and $N=49$.}
\label{aia211_lc}
\end{figure*}
\begin{figure*}[t]
\centering
\hspace{0.cm}
\includegraphics[bb=60 200 580 600,width=17.cm]{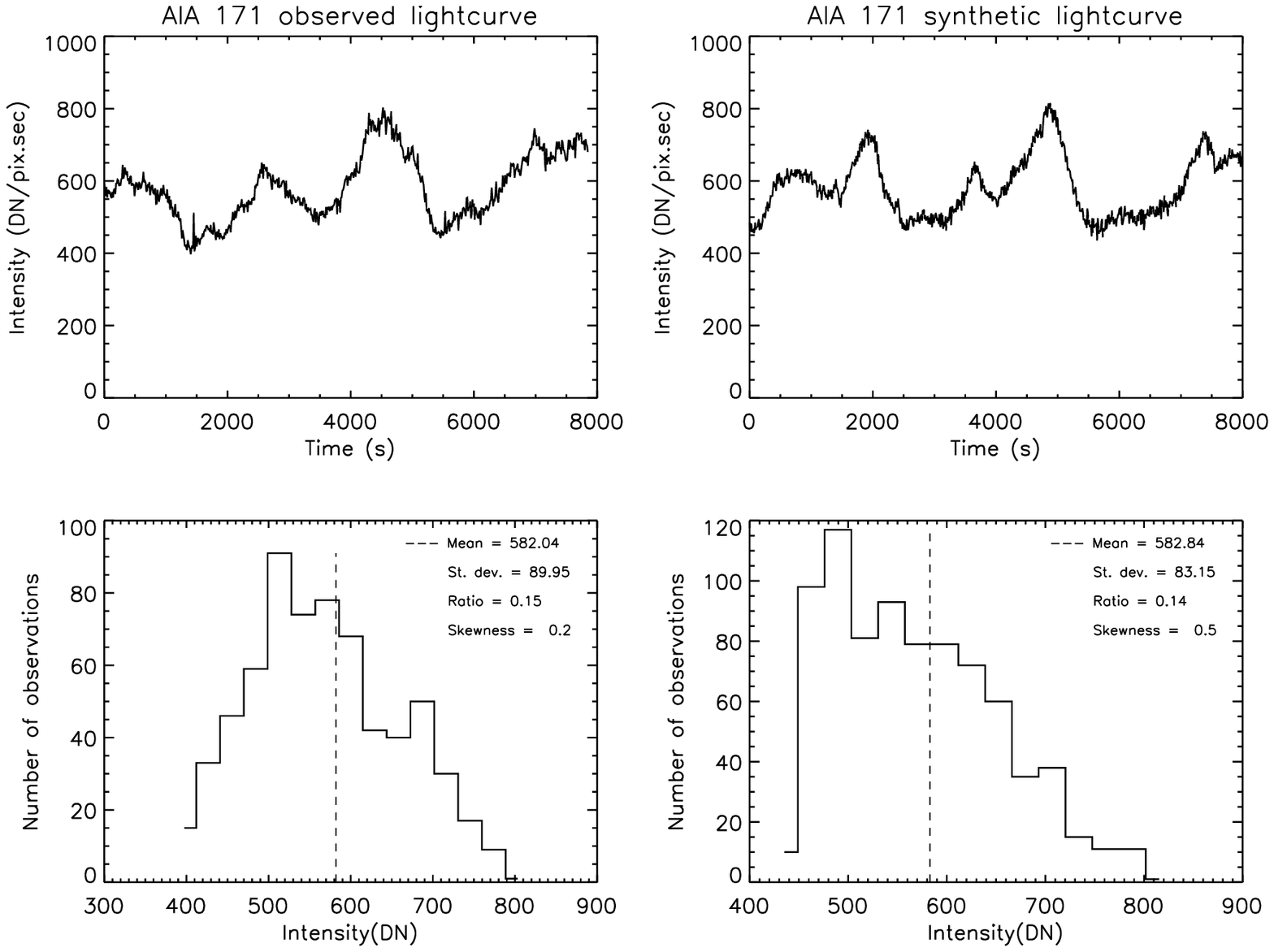}
       \caption{Idem Figure~\ref{xrt_lc} for AIA 171. The observed lightcurve corresponds to the location indicated with arrows in the corresponding panels of Figures~\ref{images} and~\ref{segments}. The parameters used for the model are $L=85$ Mm and $N=49$.}
\label{aia171_lc}
\end{figure*}
\begin{figure*}[h]
\centering
\vspace{1.5cm}
\includegraphics[bb=80 360 530 660,width=14cm]{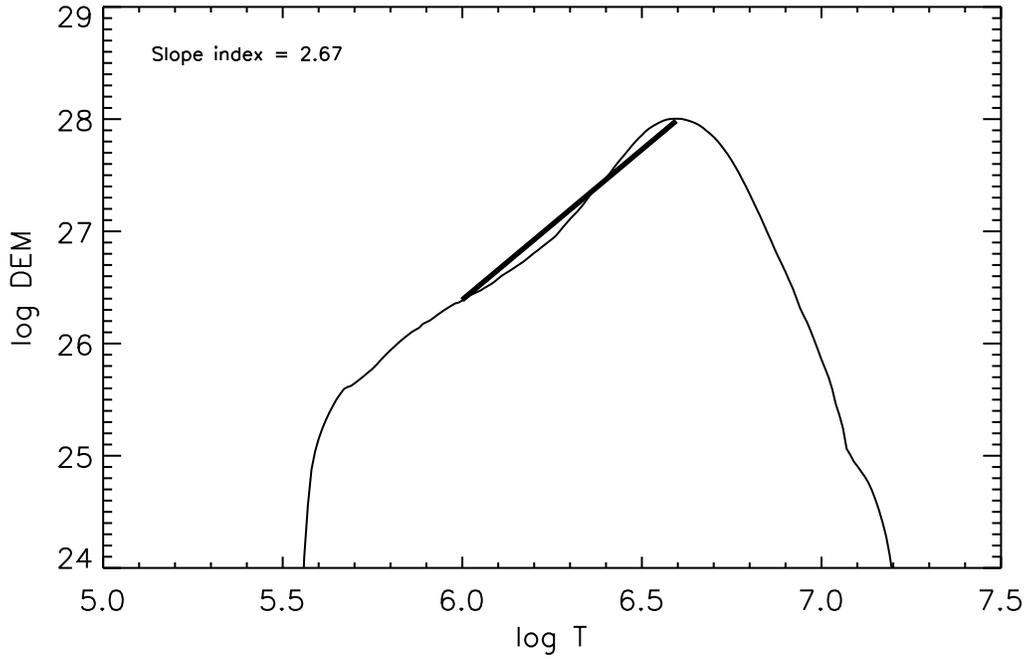}
       \caption{log(DEM) distribution as a function of log(T) for a single strand from the model. The strand length parameter is $L=$ 100 Mm. The total duration of the run is 2$\times10^5$ s with a time cadence of 1 s. The slope of the distribution is indicated by the thick segment and the corresponding index is written in the upper left.}
\label{dem}
\end{figure*}
\begin{figure*}[h]
\centering
\vspace{1.5cm}
\includegraphics[bb=80 360 530 660,width=14cm]{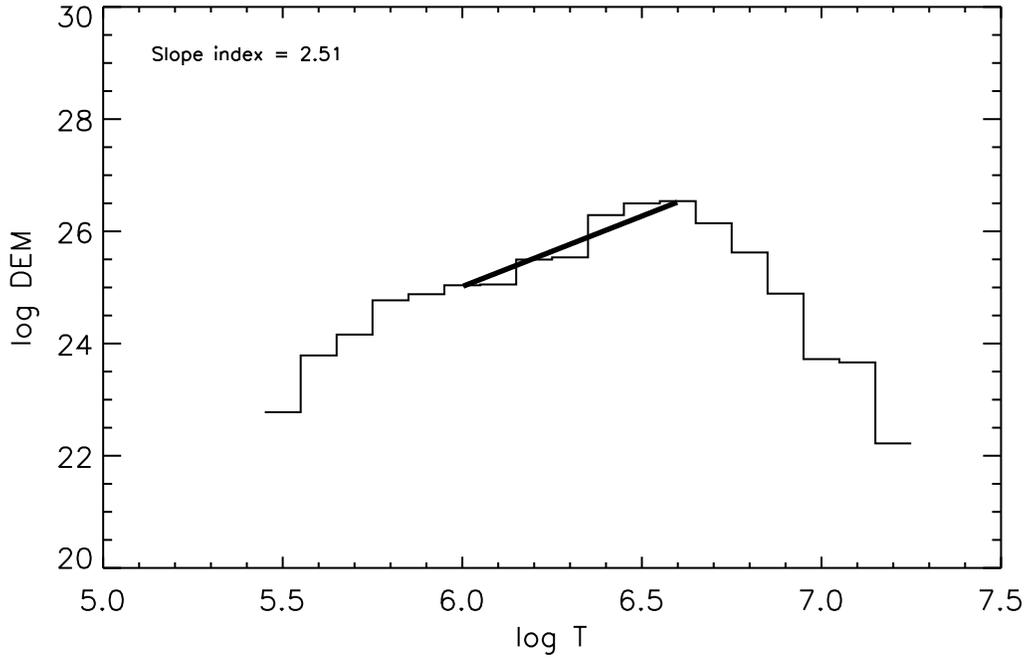}
       \caption{log(DEM) distribution as a function of log(T) for all the strands of a model with $L=$ 100 Mm, $N=$ 225. The distribution corresponds to an interval of 30 s, which is a typical Hinode/EIS integration time. The binning in log(T) = 0.1 intervals is commonly used in actual observations. The slope of the distribution is indicated by the thick segment corresponding to an index of 2.51.}
\label{obs_dem}
\end{figure*}

Using the EBTEL code we are able to compute the differential emission measure (DEM) distribution of the strands in the model. The relative amount of plasma emitting at different temperatures provides clues about how often the strands are reheated. As shown in Figures~\ref{single_strand}, nanoflares do not occur regularly in our model. The interval between successive events can correspond to high or low frequency relative to the plasma cooling time. When high frequency dominates, the strands do not reach low temperatures very often, and the cold part of the distribution is not very prominent. In that case, the slope of the log-log distribution of the DEM versus temperature is large. In the opposite case, when low frequency dominates, most of the plasma reaches lower temperatures before being reheated, and the distribution is correspondingly flatter. Whether high or low frequency dominates depends on the model parameters. The range of slopes obtained here is consistent with the observational results by other authors (Bradshaw et al. 2012, Schmelz \& Pathak 2012, Warren et al. 2012, among others).

\begin{figure*}[h]
\centering
\vspace{1.5cm}
\includegraphics[bb=90 360 530 660,width=14cm]{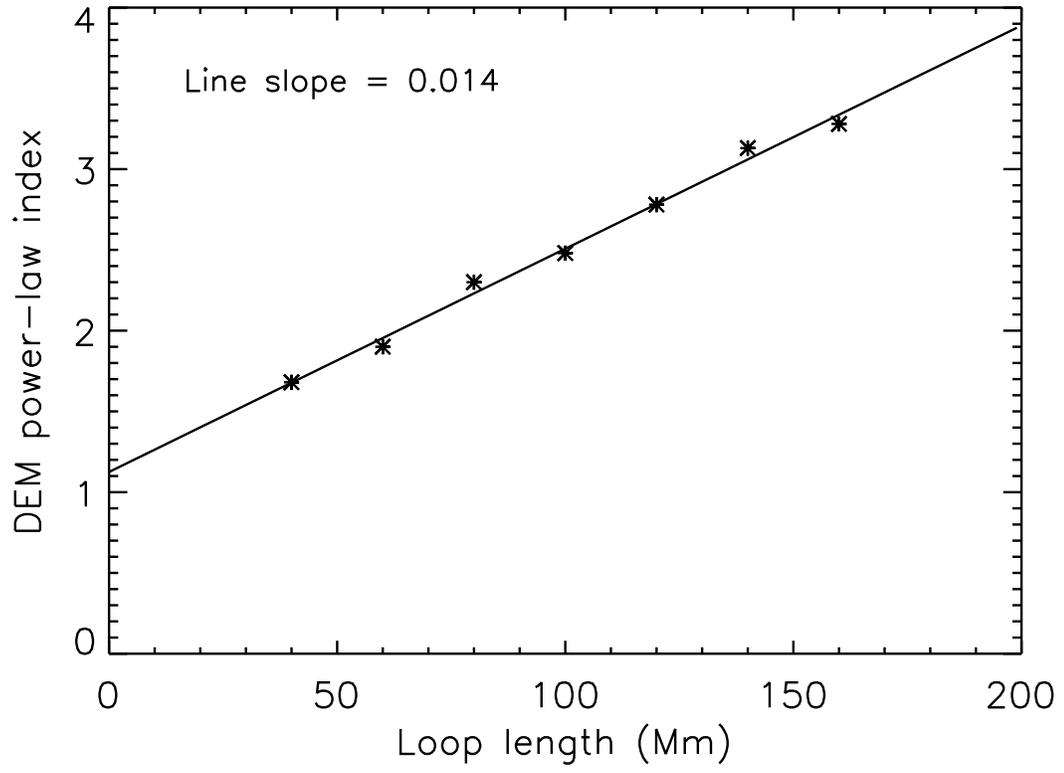}
       \caption{Scatter plot of the DEM slope index versus the loop length parameter of the model for the runs described in Section~\ref{DEM_slope}. The line corresponds to a linear regression of the plotted points, and the obtained slope is provided in the panel.}
\label{slope_length}
\end{figure*}

In Paper I we analyzed the dependence of the nanoflare frequency on the different parameters of the model. We found that in our simple CA model the frequency of the nanoflares is independent of the parameters $B_v$, $L$, and $\theta_c$ (the distance, $d$, and duration $\delta t$ of the footpoint displacement were fixed, as in this paper). However, the thermal evolution of the strands depends strongly on $L$. The initial cooling after the nanoflare is due primarily to thermal conduction, with a cooling rate that scales as $L^{-2}$. Shorter strands cool faster than longer ones. Therefore, for a given nanoflare frequency, the strands in a short loop will reach lower temperatures faster and more often than in a long loop. It is expected then, that the DEM distributions in shorter loops will have a larger contribution of plasma on the cold side and thus, smaller slopes. This is born out in our models, as shown in Figure~\ref{slope_length}. Note that this does not imply that short/long lengths produce exclusively cooling/hot strands. All the simulated loops produced by the model contain a mixture of both types. This is clearly evident in Figure~\ref{single_strand}. The key point is that there is a larger contribution of the cooling plasma (low frequency) component in shorter loops than in longer loops. The percentages of the two contributions as function of loop length were thoroughly analyzed in Paper I.


Visual inspection of Figures 1 and 2 reveals little evidence for correlation among the three observing channels, both spatially and temporally. This may seem surprising given that low frequency nanoflares should produce such a correlation. Visual impressions can be deceiving, however. Viall and Klimchuk (2012, 2013, 2015) have performed a rigorous time-lag analysis of light curves observed in different AIA channels and find widespread evidence of cooling plasma. The distribution of nanoflare frequencies is yet to be determined, however. An important point is that the cooling plasma is contained in spatially unresolved strands, as in our models, so variations in the light curves are generally very subtle. Given the lack of visual correlation in our data, we have chosen to treat the loops identified in a given channel as unique to that channel. Each loop of course produces some emission in all channels,  but a quantitative comparison is hampered by the presence of uncertain background emission. We have adopted a conservative approach and subtracted a constant background that is the minimum intensity along the vertical line in Figure 2. The actual background at any location or any time could be significantly stronger. Furthermore, the ratio of background intensities in the different channels could be variable.


One possibility regarding the apparent lack of correlation between the hot (XRT) and cooler (AIA) emissions is that the hot component has a predominance of high frequency nanoflares, but still there is some low frequency contribution such that only a small fraction of the plasma is cooled to the AIA temperatures (e.g., Fig. 4). If this low frequency component accounts for only 10\% of the hot emission (see e.g., Warren et al. 2011), its presence and any observable correlation with its evolved cool counterpart will be masked by the non-cooling 90\% high frequency contribution. Whether this component is able to explain all the cool emission observed has still to be determined. It is worth to comment that the above discussion refers only to a possible interpretation of the apparent lack of correlation between XRT and AIA observations. It does not imply that in its present form our model is able to provide these differentiated high and low frequency components. As we discussed in Paper I, the CA model has a frequency distribution which is not substantially affected by the parameters variation. When it is combined with EBTEL, the change of parameters such as the loop length and nanoflare duration affect the cooling times, so the nanoflare frequency measured with respect to the cooling time is parameter dependent. However, these changes affect all the nanoflares monolithically and are not able to produce different proportions of high and low frequency events. We hope that future, more sophisticated versions of the model, can account for multi-component nanoflare frequency distributions.


The CA-EBTEL model combination developed and studied in Paper I and applied here is a useful tool for testing the nanoflare scenario in relation to the evolution of observed loops. One of the main objectives of the present version of the model is to keep it simple to carefully study its implications. In following works, we plan to develop modified versions to account for more detailed strand interaction rules and for the evolution of loops in different parts of ARs (i.e., the core or the periphery).

\acknowledgments The authors wish to thank the anonymous referee for fruitful comments and suggestions. Hinode is a Japanese mission developed and launched by ISAS/JAXA, with NAOJ as domestic partner and NASA and STFC (UK) as international partners. It is operated by these agencies in co-operation with ESA and NSC (Norway). The AIA data used here are courtesy of SDO (NASA) and the AIA consortium. JAK's work was funded by the NASA Supporting Research and Guest Investigator programs. MLF acknowledges financial support from the Argentinean grants PICT 2012-0973 (ANPCyT), UBACyT 20020130100321 and PIP 2012-01-403 (CONICET).


\clearpage


\begin{table}
\caption{Statistical properties of comparable XRT observed and synthetic lightcurves. On each subdivision the upper row corresponds to the observed lightcurve and the lower one to the corresponding model. The first column indicates the lightcurve index (in XRT pixels) in the XRT segment evolution of the top panel of Figure~\ref{segments}. Mean, median and standard deviation intensities are in DN pix$^{-1}$ s$^{-1}$. The last column provides the model parameters used to obtain the synthetic lightcurves ($L$ is in Mm).}
\label{xrt_table}
\begin{center}
$\begin{array}{ccccccc}
$Obs. index$ & $Mean$ & $Median$ & $St. dev.$ & $Ratio$ & $Skweness$ & $Model Param.$ \\
\hline
25 & 1799.35 & 1759.11 & 310.19 & 0.17 & 0.22 &             \\
   & 1802.6  & 1773.85 & 281.10 & 0.16 & 0.63 & L=77, N=81  \\
   &         &         &        &      &      &             \\
59 & 2180.23 & 2155.02 & 214.81 & 0.10 & 0.44 &             \\
   & 2167.17 & 2148.24 & 247.52 & 0.11 & 0.40 & L=73, N=169 \\
   &         &         &        &      &      &             \\
63 & 1263.65 & 1256.63 & 150.44 & 0.12 & 0.22 &             \\
   & 1240.2  & 1233.81 & 156.06 & 0.13 & 0.35 & L=86, N=121 \\
   &         &         &        &      &      &             \\
72 & 1189.85 & 1167.68 & 156.51 & 0.13 & 0.32 & 	        \\
   & 1188.09 & 1180.72 & 172.59 & 0.14 & 0.23 & L=88, N=121  \\
   &         &         &        &      &      &             \\
90 & 716.97	 & 713.062 &  93.35 & 0.13 & 0.29 & 	        \\
   & 717.06  & 711.337 & 101.83 & 0.14 & 0.52 & L=101, N=81 \\
\hline
\end{array}$
\end{center}
\end{table}


\begin{table}
\caption{Same as Table~\ref{xrt_table} for AIA 211. Observed indices in the first column correspond to position (in AIA pixels) along the segment whose evolution is shown in Figure~\ref{segments}, middle panel.}
\vspace{0.5cm}
\label{aia211_table}
\begin{center}
$\begin{array}{ccccccc}
$Obs. index$ & $Mean$ &	$Median$ & $St. dev.$ & $Ratio$ & $Skweness$ & $Model Param.$ \\
\hline
22 & 2089.29 & 2087.61 & 147.07 & 0.07 & 0.21 & \\
   & 2091.29 & 2083.09 & 224.72 & 0.11 & 0.18 & L=47, N=441 \\
   &         &         &        &      &      &             \\
78 & 2672.27 & 2577.99 & 386.04 & 0.14 & 0.61 & \\
   & 2670.53 & 2604.71 & 416.34	& 0.15 & 0.64 & L=42, N=225 \\
   &         &         &        &      &      &             \\
130	& 1310.17 & 1283.79 & 262.71 & 0.20 & 1.03	& \\
	& 1318.69 &	1291.22	& 244.85 & 0.19 & 0.95 & L=62, N=25 \\
   &         &         &        &      &      &             \\
132	& 1563.89 & 1482.05	& 235.37 & 0.15 & 0.74 & \\
	& 1543.87 & 1524.79 & 199.63 & 0.13 & 0.59 & L=56, N=81 \\
   &         &         &        &      &      &             \\
183	& 986.5	  & 977.17 & 66.1 &	0.07 & 0.64	& \\
	& 992.32  & 990.57 & 76.2 &	0.08 & 0.25 & L=80, N=49 \\
\hline
\end{array}$
\end{center}
\end{table}


\begin{table}
\caption{Same as Tables~\ref{xrt_table} and~\ref{aia211_table} for AIA 171. Observed indices in the first column correspond to position (in AIA pixels) along the segment whose evolution is shown in Figure~\ref{segments}, bottom panel.}
\vspace{0.5cm}
\label{aia171_table}
\begin{center}
$\begin{array}{lccccccc}
$Obs. index$	& $Mean$ & $Median$	& $St. dev.$ & $Ratio$ & $Skweness$ & $Model Param.$\\
\hline
18 & 2029    &	1997.81 & 179.28 & 0.09 & 0.39 & \\
   & 2027.86 &	1978.72 & 399.43 & 0.20 & 0.56 & L=46, N=441 \\
   &         &         &        &      &      &             \\
107	& 1574.1  &	1558.97	& 125.08 & 0.08 & 0.59 &  \\
	& 1553.86 &	1537.42	& 220.38 & 0.14	& 0.51 & L=51, N=441 \\
   &         &         &        &      &      &             \\
116	& 966.53 & 956.545 & 61.7484 & 0.06 &	0.26 &	\\
	& 983.06 & 975.799 & 120.39  &	0.12 &	0.32 & L=62, N=441 \\
   &         &         &        &      &      &             \\
198	& 616.47 &	612.43  & 98.70  & 0.16 & 0.40 & \\
	& 627.53 & 613.934	& 108.28 & 0.17 & 0.70 & L=80, N=49 \\
   &         &         &        &      &      &             \\
199	& 581.79 & 575.16 & 90.30 &	0.15 & 0.27 & \\
	& 583.24 & 570.24 & 84.75 &	0.14 & 1.11 & L=85, N=49 \\
\hline
\end{array}$
\end{center}
\end{table}


\begin{table}
\caption{DEM slope indices obtained from the CA-EBTEL model for different values of the loop length, $L$. The values in each column are from the different runs at a given $L$, with the mean given at the bottom.}
\vspace{0.5cm}
\label{slope_indexes}
\begin{center}
$\begin{array}{lccccccc}
L $(Mm)$            & 40   & 60   & 80   & 100  & 120  & 140  & 160  \\
\hline
$DEM slope indexes$ & 1.59 & 1.92 & 2.03 & 2.29 & 2.73 & 3.19 & 3.08 \\
		    & 1.83 & 1.89 & 2.41 & 2.55 & 2.8  & 3.27 & 3.82 \\
                    & 1.64 & 1.81 & 2.43 & 2.25 & 3.07 & 2.69 & 3.45 \\
                    & 1.7  & 2.13 & 2.31 & 2.56 & 2.64 & 3.45 & 2.97 \\
                    & 1.63 & 1.73 & 2.33 & 2.67 & 2.68 & 3.03 & 3.07 \\
\hline
$Mean values$       & 1.68 & 1.9  & 2.30 & 2.46 & 2.78 & 3.13 & 3.28 \\
\hline
\end{array}$
\end{center}
\end{table}


\clearpage

\clearpage

\clearpage

\clearpage

\clearpage

\clearpage

\clearpage

\clearpage

\clearpage

\clearpage

\end{document}